\documentclass[12pt]{article}
\usepackage{latexsym}
\usepackage{graphics}
\usepackage{color}
\begin{document}

\begin{titlepage}
\begin{flushright}
       {\bf UK/99-21}  \\
       Dec. 1999      \\
\end{flushright}
\begin{center}

{\bf {\LARGE  Reply to Isgur's Comments on \\
\vspace*{0.3cm}
Valence QCD}}

\vspace{1.5cm}

{\bf  K.F. Liu$^{a,b}$, S.J. Dong$^b$, T. Draper$^b$, 
J. Sloan$^{b,c}$, \mbox{W. Wilcox}$^d$, and 
\mbox{R.M. Woloshyn}$^e$} \\ [0.5cm]
{\it  $^a$ SLAC, P. O. Box 4349, Stanford, CA 94309 \\
$^b$ Dept. of Physics and Astronomy,  
  Univ. of Kentucky, Lexington, KY 40506\\
$^c$ Spatial Technology, Boulder, Co \\ 
 $^d$ Dept. of Physics, Baylor Univ., Waco, TX 76798 \\
  $^e$ TRIUMF, 4004 Wesbrook Mall, Vancouver, B. C., Canada V6T 2A3}
\end{center}

\begin{abstract}
We reply to Nathan Isgur's  critique that is
directed at some of the conclusions drawn from the lattice simulation of
valence QCD, regarding the valence quark model and 
effective chiral theories.  
\bigskip
 
 
\end{abstract}

\vfill
\end{titlepage}

\section{Introduction}

     With the goal of understanding the complexity of QCD and the role
of symmetry in dynamics, we studied a field theory called Valence 
QCD (VQCD)~\cite{ldd99} in which
the Z graphs are forbidden so that the Fock space is limited to
the valence quarks. We calculated nucleon form factors, matrix elements,
and hadron masses both with this theory and with quenched QCD on a
set of lattices with the same gauge background. Comparing the results of
the lattice calculations in these two theories, we drew conclusions 
regarding the $SU(6)$ valence quark model and chiral symmetry. 
While recognizing
the goal of VQCD, Nathan Isgur disagrees on some of the conclusions 
we have drawn~\cite{isg99}.

The foremost objection raised in \cite{isg99} is to our suggestion
that the major 
part of the hyperfine splittings in baryons is due to Goldstone boson exchange
and not one-gluon-exchange (OGE) interactions. The logic of Isgur's objection
is that VQCD yields a spectroscopy vastly different from quenched QCD and
therefore the structure of the hadrons (to which hyperfine splittings in a 
quark
model are intimately tied) is also suspect so no definite conclusions are
possible.
To put this into perspective it should be emphasized at the outset that
spectroscopy is only one aspect of hadron physics examined in \cite{ldd99}. 
We have studied
the axial and scalar couplings of nucleon in terms of $F_A/D_A$ and
$F_S/D_S$, the neutron to proton magnetic moment ratio $\mu_n/\mu_p$,
and various form factors. None of these results reveal any pathologies of
hadron structure and turn out
to be close to the $SU(6)$ relations, as expected. In fact this is what 
motivated the study of valence degrees of freedom via VQCD.

In Sec. 2 we address specific issues related to spectroscopy in VQCD.
Isgur also presented more general agruments against the idea of 
boson exchange as a contributor to hyperfine effects. 
A cornerstone of his discussion is the unifying aspect of OGE in a
quark model picture. We believe that it is also natural and economical
to identify chiral symmetry as the common origin for much of the physics
being discussed here. Therefore in Sec. 3 we take the opportunity to sketch out an 
an effective theory that may serve as a framework to interpret the numerical
results of VQCD.

\section{Hadron Spectrum}

\subsection{Meson excitation --- $a_1$ -- $\rho$ mass difference}

Isgur argues that even with the `constituent quark' mass shift
incorporated into VQCD 
which lifts the baryon masses by $\sim 3 m_{const}$ and the mesons
by $\sim 2 m_{const}$, it does not restore the $a_1$ -- $ \rho$ mass splitting. 
This is a good point. However, the author's objection that the $a_1$ does not 
have an orbital excitation
energy relative to the $\rho$ is based on the non-relativistic picture
that the axial vector meson has a $p$-wave excitation as compared to the 
$s$-wave description of the vector meson. This is not necessarily 
true for the relativistic system of light quarks.  
For example, in a chirally-symmetric world, there are degenerate states 
due to parity doubling. The pion would be degenerate with
the scalar and $a_1$ would be degenerate with $\rho$. This is indeed
expected at high temperature where the chiral symmetry breaking order
parameter, $\langle \bar{\Psi}\Psi\rangle$, goes to zero.

For heavy quarks, we think VQCD should be able
to describe the vector -- axial-vector meson difference based on the 
non-relativistic picture. As seen from Figs. 25 and 28 in
Ref.~\cite{ldd99}, from $m_q a = 0.25$ on, the axial-vector meson starts to lie
higher than the vector meson. In the charmonium region ($\kappa = 0.1191$), 
we find the mass difference between them to be $502 \pm\ 80\,{\rm MeV}$. 
Indeed, this is close to the experimental difference of
$413\,{\rm MeV}$ between $\chi_{c1}$ and $J/\Psi$.

In the light quark region the near degeneracy of $a_1$ and $\rho$
is interpreted as due to the
fact that axial symmetry breaking scale, as measured by the condensates
$\langle \bar{u}u\rangle$ and $\langle \bar{v}v\rangle$, is small in
VQCD as compared to $\langle \bar{\Psi}\Psi\rangle$ in QCD~\cite{ldd99}.
As a result, there are near parity doublers in the meson spectrum. 
Note that it is consistent with the observation that dynamical mass generation,
another manifestion of spontaneously broken chiral symmetry, is also
very small in VQCD.

In the chiral theory, Weinberg's second sum rule gives the relation
$m_{a_1} = \sqrt{2} m_{\rho}$ and the improved sum rule, taking into
account of the experimental $a_1$ and $\rho$ decay constants, gives
$m_{a_1} = 1.77 m_{\rho}$~\cite{li96}. This relation is based on chiral 
symmetry, current
algebra, vector meson dominance, and the KSFR relation. These 
are based on the premise of spontaneous symmetry breaking (SSB). Otherwise,
one would expect parity doubling for $a_1$ and $\rho$. Thus, to 
explain the spectrum, we 
argue that it is sufficient to implement SSB chiral symmetry, not
necessarily the $p$-wave orbital excitation as in the non-relativistic
theory. In other words, by restoring the spontaneously broken 
$SU(3)_L \times SU(3)_R \times U_A(1)$ symmetry 
to VQCD which has only $U_q(6)\times U_{\bar{q}}(6)$, it is possible 
to restore the physical mass difference between $a_1$ and $\rho$
to be consistent with Weinberg's sum rule.

\subsection{Hyperfine splittings}

As for hyperfine splittings,
we have argued that the one-gluon-exchange 
is not the major source since OGE is still contained in VQCD.
Being magnetic in origin, the color-spin interaction is related
to the hopping of the quarks in the gauge background in the spatial
direction~\cite{ef81}. VQCD does not change this from QCD; 
the $\vec{\sigma}\cdot \vec{B}$ term is present in the
Pauli spinor representation of the VQCD action.
Thus, we are forced to draw the conclusion that one-gluon-exchange
type of color-spin interaction, i.e. $\lambda^c_i \cdot \lambda^c_j
\vec{\sigma}_i \cdot \vec{\sigma}_j$, cannot be responsible for the
majority part of the hyperfine splittings between $N$ and $\Delta$ and
between $\rho$ and $\pi$. While we suggested that the Goldstone
boson exchange is consistent with the Z-graphs and maybe responsible 
for the missing hyperfine interaction in the baryons (Fig. 1), it is 
correctly pointed out by Isgur that there is no such $q\bar{q}$
exchange between the quark and anti-quark in the meson.

\begin{figure}[h]
\begin{center}
\includegraphics{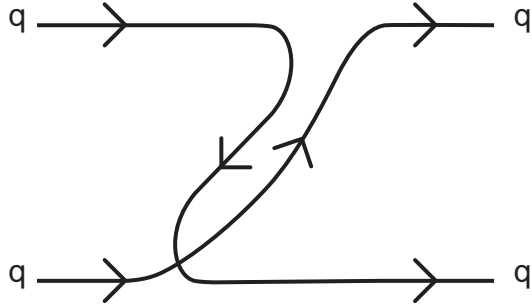}
\end{center}
\caption{ Z-graph between two quarks in a baryon.}
\end{figure}

One therefore has to consider the
possibility that the hyperfine splitting mechanism in the light quark sector
is different in mesons from that in the baryons.
The numerical results of QCD and VQCD do not, by themselves, reveal the 
interaction mechanism. A mapping to some model is necessary to make an
interpretation.
We consider the $SU(3)$ Nambu-Jona-Lasinio 
NJL model as an example.
Starting with a color current-current coupling~\cite{vw91} 
\begin{equation}   \label{col-cur}
- 9/8 G (\bar{\psi}t^a\gamma_{\mu}\psi)^2,
\end{equation}
 it is convenient to
consider Fierz transform to include the exchange terms. The Lagrangian
for the color-singlet $q\bar{q}$ meson then takes the following
$SU(3)_L \otimes SU(3)_R$ symmetric form with dimension-6 operators
for the interaction
\begin{eqnarray}
\lefteqn{
{\cal L}_{NJL} = \bar{\psi}(i \not{\partial} - m_0)\psi +
 G \sum_i [(\bar{\psi}\frac{\lambda_i}{2}\psi)^2
+(\bar{\psi}\frac{\lambda_i}{2}i\,\gamma_5\psi)^2] 
} \nonumber \\
& & \phantom{{\cal L}_{NJL} = }
- G/2 \sum_i [(\bar{\psi}\frac{\lambda_i}{2}\gamma_{\mu}\psi)^2
+(\bar{\psi}\frac{\lambda_i}{2}\gamma_{\mu}\gamma_5\psi)^2].
\end{eqnarray}
The scalar four-fermion interaction can generate a dynamical quark mass
\begin{equation}
m_d = G \langle \bar{\psi}\psi\rangle.
\end{equation}  
in the mean-field approximation. This is illustrated in
Fig. 2. While all the meson masses are lifted up by the dynamical
quark masses, the attractive pseudo-scalar four-fermion interaction
brings the pion mass back to zero making it a Goldstone boson.
The repulsive vector and axial-vector four-fermi interaction makes
the $\rho$, at $\sim 770\,{\rm MeV}$, slightly higher than twice $m_d
= 360\,{\rm MeV}$. Similarly, the $a_1$ mass is calculated at   
$m_{a_1} \simeq 1.2\,{\rm GeV}$, which is not far from the Weinberg's 
sum rule relation 
$m_{a_1} = \sqrt{2}m_{\rho}$. We see that with one parameter, $G$, 
the meson masses can be reasonably described in the NJL model without
the $q\bar{q}$ type of meson exchange as in Fig. 1. In addition, 
current algebra relations such as the Gell-Mann-Oakes-Renner relation
\begin{equation}
m_{\pi}^2 f_{\pi}^2 = - \frac{m^0_u + m^0_d}{2} \langle \bar{u}u +
\bar{d}d \rangle,
\end{equation}
are satisfied. The crucial ingredient here is spontaneous chiral symmetry
breaking which is characterized by non-vanishing $f_{\pi}$ and quark
condensate $\langle \bar{\Psi}\Psi\rangle$, and the existence of
Goldstone bosons.

\begin{figure}[ht]
\begin{center}
\includegraphics{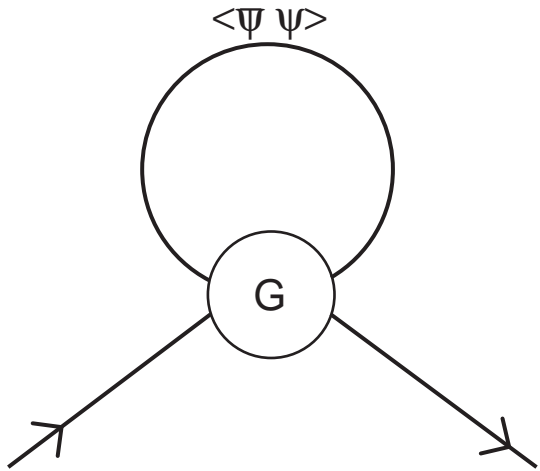}
\end{center}
\caption{The dynamical mass is generated through the four-fermi
interaction with a mean-field approximation.}
\end{figure}

    We should point out that although the color current--current coupling in 
Eq. (\ref{col-cur}) is reminiscent of the one-gluon-exchange interaction
with the $q^2$ in the gluon propagator replaced by a cut-off $\Lambda^2$ 
which reflects the short-range nature of the interaction,
it is the covariant form for relativistic quarks not the 
one-gluon exchange potential in the non-relativistic reduction. It is 
the latter which has
been considered as the standard form for hyperfine and fine splittings
in the valence quark model.

   As illustrated through the NJL model, it is possible to have different
mechanisms for hyperfine splitting in the baryons and mesons. 
In the baryons, the hyperfine splitting {\em can} be largely due to the meson 
exchanges between the quarks in the $t$-channel (Fig. 1); whereas in 
the mesons, 
it is the $s$-channel short-range four-fermion coupling (Fig. 3) that 
give rise to the hyperfine splittings. Although they appear to be
different mechanisms, both of them are based on spontaneously broken   
chiral symmetry.

\begin{center}
\begin{figure}[ht]
\includegraphics{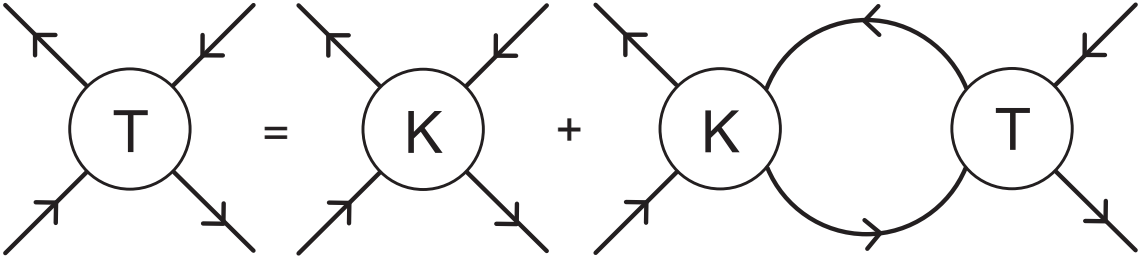}
\caption{Bethe-Salpeter equation for the meson T-matrix.}
\end{figure}
\end{center}

    The author displayed the spectrum ranging from heavy--heavy mesons 
($b\bar{b}, c\bar{c}$) to light--light mesons ($s\bar{s}$ and
isovector light quarkonia) in Fig. 4 of his paper~\cite{isg99} 
which suggests a smooth
trend as a function of the quark mass and argues for a universal
OGE hyperfine interaction with a strength proportional to $1/m_Q^2$.
We have pointed out in our VQCD paper~\cite{ldd99} from the
outset that we believe the heavy--heavy mesons are well described
by a non-relativistic potential model including the OGE; this is  
supported by the lattice calculations~\cite{slo95,shi97,ses98,dav98}.
It is the validity of OGE in the light--light mesons sector that we question.
What have been neglected in Fig. 4 of Ref.~\cite{isg99} are the
$1^{++}$ and $0^{++}$ mesons. Had these been put in, one would have seen
that $a_0(1430)$ lies higher than $a_1(1260)$ and $a_2(1320)$. This
ordering between $1^{++}$ and $0^{++}$ mesons is reversed from 
that in the charmonium family where $\chi_{c1}(3510)$ lies higher than
$\chi_{c0}(3415)$. There is an indication from the lattice calculation
that this cross-over occurs at about the strange mass region~\cite{
ko98}. As far as we know, this pattern of order reversal in the
fine splitting as the quark mass becomes light  
cannot be accommodated in the OGE picture.

Also shown in Fig. 5(a) of Ref.~\cite{isg99} are the hyperfine
splittings of the ground state heavy-light mesons. We concur that
the splittings of $B^*(5325)$ -- $B(5279)$ and $D^*(2010)$ -- $D(1869)$
are quite consistent with the matrix elements of the hyperfine
interaction $\vec{\sigma}_Q \cdot \vec{B}/2m_Q$ and that it clearly 
demonstrates the $1/m_Q$ behavior of the heavy quark. We never 
questioned the relativistic corrections of the heavy quarks. It
is with light quarks that we think OGE has problems.
For example, consider
the similar splittings for the heavy--light mesons with different 
light quarks. The mass difference between
$D^*(2010)$ and $D(1869)$ is $140.64 \pm 0.10\,{\rm MeV}$. This is 
practically the same as that
between $D^*_s(2110)$ and $D_s(1969)$ which is $143.9 \pm 0.4\,{\rm MeV}$.
There is no indication of
the $1/m_q$ dependence on the light quark mass as required by the
OGE potential. Similarly we find that $m_{B^*} - m_B = 45.78 \pm
0.35\,{\rm MeV}$ is identical to $m_{B^*_s} - m_{B_s} = 47.0
\pm 2.6\,{\rm MeV}$. Again, there is no $1/m_q$ dependence.

\section{Effective Theory for Both Mesons and Baryons} 

Besides commenting on the spectroscopy specific to VQCD, 
Isgur also questioned 
the meson exchange picture on more general grounds.
Since this issue has been raised, we take the opportunity to 
extend our discussion although it is outside the scope of VQCD.

Perhaps the most serious challenge to the meson exchange picture
in the baryons is the possibility of meson exchanges between the 
quark and anti-quark in the iso-singlet meson. It is pointed out
by Isgur that the annihilation diagram  depicted in
Fig. 6 in Ref.~\cite{isg99} in terms of the quark lines is OZI 
suppressed in QCD. We should add that it is $O(1/N_c^2)$ suppressed
as compared to one-pion-exchange between the quark pairs in the
baryon (Fig. 1) in the large $N_c$ analysis. On the other hand,
interpreting this as a Goldstone boson exchange between the quark 
and anti-quark in
the iso-singlet mesons, such as a kaon exchange,  leads to  large
$\omega - \phi$ mixing. How does one reconcile the apparent 
contradiction? The short answer is that there is no such process
in the effective theory of mesons. It is inconsistent, within the
renormalization group approach to effective theories, to consider this QCD
annihilation process as a meson exchange between the quark and anti-quark in
the meson. To see this, we shall use the NJL model as
an illustration.

\subsection{Bosonization}

We shall follow the example given by U. Vogl and W. Weise~\cite{vw91}
for a simple $U(1)_V \otimes U(1)_A$ symmetric Lagrangian
\begin{equation}
{\cal L} = \bar{\psi}(i \not{\partial} -m_0)\psi + 
G [(\bar{\psi}\psi)^2
+(\bar{\psi}i\,\gamma_5\psi)^2].
\end{equation}
To bosonize this theory, one needs to integrate out the fermions. One
can follow the Hubbard-Stratonovich transformation~\cite{hs59} by introducing
Gaussian auxiliary boson fields $\sigma$ and $\pi$ with the
Lagrangian $ - \mu^2/2 (\sigma^2 + \pi^2)$ and the partition
function becomes
\begin{equation}  \label{aux}
{\cal Z} = {\cal N}\int {\cal D}\sigma {\cal D}\pi {\cal D}\bar{\psi}
{\cal D}\psi\, e^{i \int d^4 x \bar{\psi}[i\,\not{\partial} - m_0
- \mu \sqrt{2G} (\sigma + i \gamma_5 \pi)]\psi - \mu^2/2 (\sigma^2 
+ \pi^2)},
\end{equation}
after a linear shift of the fields $\sigma$ and $\pi$. Note here,
the $\sigma$ and $\pi$ are the auxiliary fields with no kinetic
terms.

   At this stage, one can integrate the fermion field with the
quadratic action to obtain the fermion determinant. This gives an
effective action with the ${\rm tr} \ln M$ Lagrangian, where $M$ is
the inverse quark propagator between the square brackets in
Eq. (\ref{aux}). Expanding the ${\rm tr} \ln M$ to the  second order in
the derivative $\partial_{\mu}$ for the low energy long wavelength 
approximation, the effective Lagrangian becomes
\begin{eqnarray}   \label{bos}
\lefteqn{ 
{\cal L}_{eff}(\sigma, \pi) =  \frac{1}{2}[(\partial_{\mu} \sigma)^2
+ (\partial_{\mu} \pi)^2] - \frac{1}{2}m_{\pi}^2 \pi^2 - \frac{1}{2}
m_{\sigma}^2 \sigma^2 } \nonumber \\
& & \phantom{{\cal L}_{eff}(\sigma, \pi) = }
- \frac{2 m^2}{f_{\pi}} \sigma
(\sigma^2 + \pi^2) - \frac{ m^2}{2 f_{\pi}} \pi (\sigma^2 + \pi^2)^2, 
\end{eqnarray}
where $m = m_0 + \mu\sqrt{2 G}\langle \sigma \rangle =
m_0 - 2G \langle \bar{\Psi}\Psi\rangle$.
Besides giving $\pi$ and $\sigma$ masses as the physical mesons, it
also gives the explicit meson-meson couplings.

Thus, to construct an effective theory below the meson confinement scale,
which corresponds to the chiral symmetry breaking scale 
$\Lambda_{\chi} = 4 \pi f_{\pi} \simeq 1\,{\rm GeV}$ as we shall see later, 
one can take the
following equivalent approaches: In the first one, one can
introduce higher dimensional operators like $(\bar{\psi}\psi)^2,
(\bar{\psi}i\,\gamma_5\psi)^2,(\bar{\psi}\gamma_{\mu}\psi)^2,
 (\bar{\psi}\gamma_{\mu}\gamma_5\psi)^2$ to the usual QCD
Lagrangian and tune the couplings to match to QCD above
$\Lambda_{\chi}$. Many improved lattice actions are constructed
this way in order to do numerical simulation at a lower lattice 
cut-off or larger lattice spacing in order to save computer time
~\cite{has98}. In the second approach, one can introduce auxiliary
fields $\pi, \sigma, \rho, a_1$, etc. to replace the four-fermion
operators with couplings to fermion bilinears and multi-auxiliary-field
couplings as in Eq. (\ref{aux}). This form has been considered in
lattice QCD simulations~\cite{bst95,kls98} to control the singular
nature of the massless Dirac operator. The third approach is to
bosonize the theory by integrating out the fermion fields and performing
derivative expansion of the ${\rm tr} \ln M$ action from the fermion loop 
as in Eq. (\ref{bos}).
An extensive and successful model of this kind has been
developed~\cite{li95} where $\rho$ is predicted to be close to the
experimental value and $a_1$ mass is related to the $\rho$ via
the modified Weinberg sum rule~\cite{li96}. VMD and the KSFR relation
are satisfied. In addition, the pion form factor, $\pi \pi$ scattering,
and a host of meson decays are all in good agreement with the
experiments.   

We see that in none of the above three equivalent approaches is there a 
coupling between the quark and physical mesons. Thus, there is no OPE
between the quark-anti-quark pair in the meson. Since one is
below the meson confinement scale $\Lambda_{\chi}$, the meson 
fields are the relevant degrees of freedom. Once one integrates
out the fermion fields in the meson in favor of the physical
meson fields, it would be inconsistent to construct a meson model with
couplings between quarks and physical mesons. Of course, this
does not preclude short-range couplings between $u\bar{u}$,
$d\bar{d}$ and $s\bar{s}$ in the $s$-channel to resolve the $U_A(1)$ anomaly 
and give $\eta'$ a large mass via the contact term of the topological
susceptibility~\cite{wv79}.  

Then how does one justify the $\sigma$ - quark model that one
proposes as an effective theory for the baryons? To realize this one has
to make a distinction between the meson and the baryon.

\subsection{Chiral effective theory for baryons}

   In view of the observation that mesons have form factors in the
monopole form and baryons have form factors in the dipole form,
the $\pi NN$ form factor is much softer than the $\rho \pi \pi$
form factor, we suggest that the confinement scale of quarks in the
baryon $l_B$ is larger than $l_M$ -- the confinement scale between the 
quark and anti-quark in the meson; that is,
\begin{equation}  \label{scale}
l_B > l_M.
\end{equation} 
This is consistent with the large $N_c$ approach where the mesons 
are treated as point-like fields and the baryons emerge as solitons
with a size of order unity in $N_c$.
Taking the $l_M$ from the $\rho \pi \pi$ form factor gives
$l_M \sim 0.2\,{\rm fm}$. This is very close to the chiral symmetry
breaking scale set by $\Lambda_{\chi} = 4 \pi f_{\pi}$. We consider
them to be the same, i.e. below $\Lambda_{\chi}$, operators of 
mesons fields become relevant operators. As for the baryon confinement
scale, we take it to be the size charactering the meson-baryon-baryon
form factors. Defining the meson-baryon-baryon form factors from
taking out the respective meson poles in the nucleon pseudoscalar, 
vector, and axial form factors (see Fig. 17 in Ref.~\cite{ldd99}),
we obtain $l_B \sim 0.6 - 0.7\,{\rm fm}$. This satisfies the inequality in
Eq. (\ref{scale}). Thus, in between these two scales $l_M$ and
$l_B$, one could have coexistence of mesons and quarks in a baryon.

We give an outline to show how to construct a chiral effective theory for
baryons. In the intermediate length scale between $l_M$ and $l_B$,
one needs to separate the fermion field into a long-range one and
a short-range one
\begin{equation}
\psi = \psi_L + \psi_S,
\end{equation}
where $\psi_L/\psi_S$ represent the infrared/ultraviolet part of the
quark field with momentum 
components below/above $1/l_M$ or $\Lambda_{\chi}$. We add to the
ordinary QCD Lagrangian irrelevant higher dimension operators 
with coupling between bilinear quark fields and auxiliary fields as
given in Ref.~\cite{li95}. However, we interpret these quark fields 
as the short-range ones, i.e. $\psi_S$ and $\bar{\psi}_S$.
Following the procedure in Ref.~\cite{li95}, one can integrate out the
 $\psi_S$ and $\bar{\psi}_S$ fields and perform the derivative expansion to
bosonize the short-range part of the quark fields. This leads to
the Lagrangian with the following generic form:
\begin{equation}  \label{ect}
{\cal L}_{\chi QCD} = {\cal L}_{QCD'}(\bar{\psi}_L, \psi_L, A_{\mu}^L)
+ {\cal L}_M(\pi, \sigma, \rho, a_1, G, ...) + {\cal L}_{\sigma q}
(\bar{\psi}_L, \psi_L, \pi, \sigma, \rho, a_1, G, ...). 
\end{equation}
${\cal L}_{QCD'}$ includes the original form of QCD but in terms of 
the quark fields $\bar{\psi}_L, \psi_L$, and
the long-range gauge field $ A_{\mu}^L$ with renormalized couplings; it 
also includes higher-order covariant derivatives~\cite{war88}. 
${\cal L}_M$ is the meson effective
Lagrangian, e.g. the one derived by Li~\cite{li95} which should include the
glueball field $G$. Finally, ${\cal L}_{\sigma q}$ gives the coupling
between the $\bar{\psi}_L, \psi_L$, and mesons. As we see, in this
intermediate scale, the quarks, gluons, and mesons coexist and
meson fields {\em do\/} couple to the quark fields, but it is $\psi_L$ that
the mesons couple to, {\em not\/} $\psi_S$.
Going further down below the baryon confinement scale $1/l_B$, one
can integrate out $\bar{\psi}_L, \psi_L$ and $A_{\mu}^L$, resulting in
an effective Lagrangian ${\cal L}(\bar{\Psi}_B, \Psi_B, \pi, \sigma, \rho, 
a_1, G, ...)$ in terms of the baryon and meson fields~\cite{wkw99}. This would 
correspond to an effective theory in the chiral perturbation theory.

\begin{figure}[ht]
\begin{center}
\includegraphics{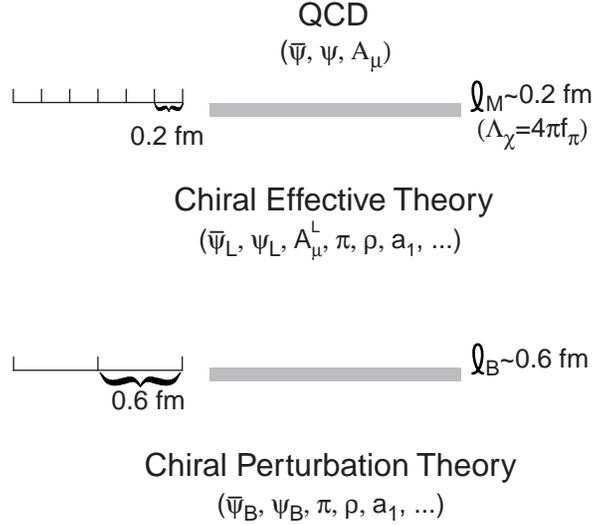}
\end{center}
\caption{A schematic illustration of the the two-scale delineation of
the effective theories. The shaded bars mark the positions of the 
cutoff scales $l_M$ and $l_B$  separating
different effective theories.}
\end{figure}

Fig. 4 is a schematic illustration of effective theories partitioned
by the two scales of $l_M$ and $l_B$. We should point out that although
we adopt two scales here, they are distinct from those of
Manohar and Georgi~\cite{mg84}. In the latter, the $\sigma$ -- quark model
does not make a distinction between the baryons and mesons. As such,
there is an ambiguity of double counting of mesons and
$q\bar{q}$ states. By making the quark-quark confinement length scale
$l_B$ larger than the quark--anti-quark confinement length scale $l_M$, one
does not have this ambiguity. The outline we give here is a systematic
way of constructing the effective theory at appropriate scales following
Wilson's renormalization group approach~\cite{wil74,pol84}. 

We see from Fig. 5 that the ${\cal L}_{\sigma q}$ part of the effective
chiral theory in Eq. (\ref{ect}) is capable of depicting meson
dominance (Fig. 5(a)), the quark Z-graphs and cloud degree of freedom via
the meson exchange current (Fig. 5(b)), and the sea quarks in the disconnected 
insertion via the meson loop (Fig. 5(c)) in a baryon. These correspond to 
the dynamical quark 
degrees of freedom in QCD as we alluded to in the study of baryon form 
factors in the path-integral formulation~\cite{ldd99}. 
On the other hand, when one considers the chiral perturbation theory at 
energy lower than $1/l_B \sim 300\,{\rm MeV}$,
the dressing of baryons with meson clouds
(Fig. 6) no longer distinguishes the cloud-quarks from the sea-quarks.

\begin{figure}[ht]
\includegraphics{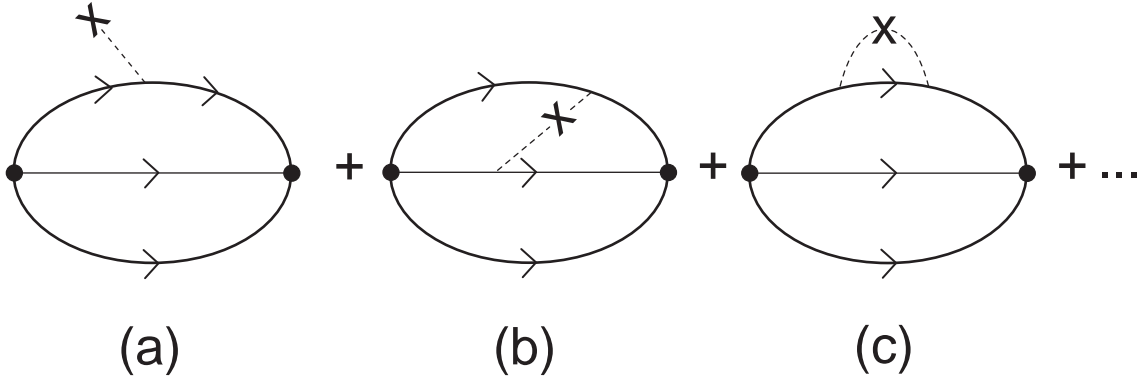}
\caption{The $\sigma$-quark model description of (a) meson dominance,
(b) cloud quarks via meson exchange current, and (c) sea quarks via
the meson loop.}
\end{figure}
\vspace{3cm}
\begin{figure}[ht]
\includegraphics{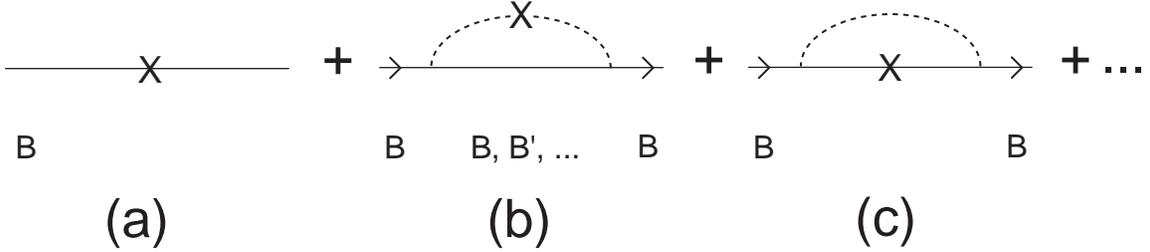}
\caption{(a) Direct baryon contribution and (b) \& (c) meson loop
contribution in the chiral perturbation theory.}  
\end{figure}

One important aspect of constructing effective theories based on
the renormalization group is that chiral symmetry and other symmetries of
the theory should be preserved 
as one changes the cut-off so as to ensure universality.

As we see from the above construction of effective chiral theories, there is
no large OZI-violating meson exchange between the quark and anti-quark in an 
iso-singlet meson. The problem that Isgur perceives for the
meson exchange in the iso-singlet meson is simply not there.

\section{Conclusions}

   As stressed at the beginning, hadron spectroscopy is only
one of the many facets of hadron physics. At low energies, there
is a lot of evidence that chiral symmetry is playing a crucial role,
for example, in the $\pi \pi$ scattering, the Goldberger-Treiman relation, 
the Gell-Mann-Oakes-Renner
relation, the Kroll-Ruderman relation, the KSRF relation, and
Weinberg sum rules.

As far as light hadrons are concerned, it is natural to expect 
chiral symmetry to play a role in spectroscopy also.
For many years, various chiral models have been successful in 
describing the pattern of masses in the meson sector in addition to
scattering and decays. Now it appears that the chiral quark picture 
can give a reasonable explanation of the baryon spectroscopy as well
as structure.

Finally, 
we echo Isgur's comment `while qQCD describes both the $\rho - \pi$ and
$\Delta - N$ splittings, they are both poorly described in vQCD. 
It would be natural and economical to identify a common origin for
these problems.' It is proposed  that chiral symmetry is this common
origin, albeit it may have different dynamical realization in mesons
and baryons. We suggest it is chiral symmetry that is the essential physics
multilated in VQCD and that this is manifested by the suppression of dynamical
mass generation, approximate parity doublets, 
the incorrect $U(6)$ symmetry and the disappearance of hyperfine
splittings.  We expect that effective chiral theories or models that
incorporate the spontaneously broken
$SU(3)_L \times SU(3)_R \times U_A(1)$ symmetry will have
the relevant dynamical degrees of freedom necessary to delineate the 
structure and spectroscopy of both mesons and
baryons of light quarks at a scale below $\sim 1\,{\rm GeV}$.

\section{Acknowledgment}
 
We thank S. Brodsky and B. A. Li for illuminating discussions. We thank
M. Peskin for pointing out Ref.~\cite{war88} to us.
This work is partially supported by U.S. DOE grant No. DE-FG05-84ER40154
and NSF grant No. 9722073.

\end{document}